# Curling Superconducting Vortices by Nematic Fluxon-Skyrmion Interaction

Xiaoqing Bao[1,2,†], Shiyuan Wang[1,2,†], Dekun Shen[1,2,†], Guibin Lan[3], Tianqi Wang[1,2], Yuelin Zhang[1,2], Guoqiang Yu[3], Ka Shen[1,2], Haiwen Liu[1,2,*] and Jinxing Zhang[1,2,*]

[1]*School of Physics and Astronomy, Beijing Normal University, Beijing 100875, China.*
[2]*Key Laboratory of Multiscale Spin Physics, Ministry of Education, Beijing 100875, China*
[3]*Beijing National Laboratory for Condensed Matter Physics, Institute of Physics, Chinese Academy of Sciences, Beijing 100190, China.*

†These authors contributed equally to this work.
*Correspondence to: haiwen.liu@bnu.edu.cn; jxzhang@bnu.edu.cn

## Abstract

Superconducting vortex represents a topological entity in type-II superconductors, where the prototypical manifestation of such a macroscopic quantum state holds great promise for applications in future quantum technologies. Deterministic control of the vortex dynamics at the nanoscale is central to enhance superconductors' critical thresholds and design next-generation superconducting quantum devices. However, it is still scientifically challenging due to the intrinsic nature of the vortex pinning landscapes in superconductors. Here, we observe an even-in-field Hall-like signal with about two order of magnitude enhancement in a type-II superconductor/chiral magnet heterostructure, which confines to the field-temperature window where the flux flow coexists with magnetic skyrmions. The polarity of such a giant even-in-field transverse response reverses between orthogonal current directions, suggesting an on-track vortex flow driven by a nematic fluxon−skyrmion interaction analyzed via the extended vortex-viscosity tensor. More critically, a deterministic manipulation of skyrmions at the nanoscale within the hybrid system enables selective and reversible controls over the dissipative behavior and therefore the reconfigurable fluxon transport, mimic the precise control of curling stone in winter Olympic, thereby opening a new avenue towards the exploration of programmable topological and superconducting quantum devices.

In type-II superconductors, a superconducting vortex is a quantized filament comprising a normal-state core and surrounding supercurrents[1,2]. Traditionally engineered pinning landscapes have long been used to suppress vortex motion, mitigate dissipation, and enhance critical superconducting thresholds[3,4]. Today, precise and reversible manipulation of vortex is a prerequisite for developing vortex-based quantum technologies[5-9], with potential applications in vortex-based memory cells[6], quantum sensing[10], topological quantum computing[7,8], etc. Vortex dynamics reflect a force balance among the current-induced driving force, the transverse Magnus force, and the viscous drag force, where the vortex velocity deviates only slightly from the direction of Lorentz force in conventional superconductors (Magnus coefficient is much less than vortex viscosity)[11,12], as shown in Fig. 1a. This drag is inherently dissipative in superconductors, while its microscopic mechanism remains heavily debated over past 60 years[11], spanning Bardeen-Stephen quasiparticle currents[12], Anderson phase slips[13,14], and spectral flow models[15,16]. Steering vortex at the nanoscale remains formidable because conventional pinning landscapes[17] and the fixed interfacial symmetry[18] are established after material growth or device fabrication, so that the vortex cannot be reconfigured during operations[19]. Thus, realizing efficient and programmable fluxonic devices requires a paradigm shift to actively rewritable landscapes capable of spatial and reversible modulation of the vortex viscosity[17,20].

An effective route to reshape vortex dynamic may be achieved by constructing a coupling of vortex with other particles[10] or quasiparticles[21]: phonon[22], magnon[23] or photon[24], even including topological quasiparticles such as skyrmions[25] etc., which have enabled to probe and reconfigure vortex as an active elements in hybrid quantum systems[10,21-26]. Among them, magnetic skyrmions are topologically stable quasiparticle with chiral and gradient spin textures that can be deterministically created, moved and annihilated at nanoscale[27-32]. Over the past decade, the interplay between these topological magnetic and superconducting entities has inspired predictions of emergent static composite states[33-36], notably topological superconductivity[37] and Majorana zero modes[38,39], etc. Beyond those static ones, the nanoscale spin gradient of a skyrmion could play critical roles in their dynamics[36,40]: it may experience an unforeseeable exchange-mediated drag channel when a superconducting vortex moving across a magnetic skyrmion, as illustrated in Fig. 1b, offering both a fresh window on the dissipation landscape and a rewritable knob for controlling vortex motion. However, despite quite a few studies attempt to locally probe the interplay of skyrmions and vortices[25,33-36,40,41], the effective control of the Abrikosov vortex (or fluxon) transport and emergent quantum phenomena has been largely hindered by the limitation of large-area stabilized skyrmion phase within wide windows of temperature and magnetic field[42].

In this work, we construct an artificial hybrid quantum heterostructure consisted of a strongly correlated half-metallic ferromagnet $La_{0.67}Sr_{0.33}MnO_3$ (LSMO) and a type-II superconducting Nb thin film. This ferromagnet provides a brand-new platform to hold a large-area stable skyrmion (SkX) phase within wide ranges of temperature/magnetic field[43]. In the superconducting transition regime of this hybrid system, we observe a two order of magnitude enhancement of an even-in-field Hall-like signal. The reversed polarity of this giant transverse response between orthogonal current directions provides a direct signature of a nematic vortex-skyrmion interaction. Ultimately, we demonstrate the precise manipulation of individual skyrmions, thereby affording selective and reversible control over the vortex dynamics at nanoscale.

## An even transverse response in the type-II superconductor-chiral magnet hybrid system

To investigate the dynamics of superconducting vortex in the hybrid system, we fabricated the Hall bars from a Nb (70 nm) film and Nb (70 nm)/LSMO bilayers. A large-area skyrmion phase over a broad range of temperature and magnetic field can be stabilized in the 150-nm LSMO (SkX) layer[43]. In comparison, we prepared identical Hall bars from a Nb/LSMO bilayer with a 15-nm LSMO without skyrmions (non-SkX)[43]. The superconducting behaviors of the above samples can be seen in Fig. S2, indicating the nature of high-quality superconducting layers in the hybrid system. Details of the thin-film growth, device fabrication, domain imaging, and transport measurements are provided in Supplementary Section 1. Structural and room-temperature magnetic characterization are shown in Supplementary Fig. S1 and Section 2, while superconducting transitions and transport in Nb/non-SkX are provided in Supplementary Figs. S2 and S3 and Section 3 and Section 4.

The longitudinal ($\rho_{xx}$) and transverse ($\rho_{xy}$) resistivities of the devices were measured at 7.6 K, as schematically shown in Figs. 2a and 2b. Distinct from the $\rho_{xx}$ and $\rho_{xy}$ in the bare Nb film (Fig. 2c), we observe a giant Hall-like signal exhibiting the same polarity for both positive and negative magnetic fields within the superconducting transition regime of the Nb/SkX hybrid system as seen in Fig. 2d. Such an even-in-field transverse response represents a symmetry forbidden in conventional Hall effect. Crucially, this anomalous signal is entirely absent in both the bare Nb and Nb/non-SkX samples (grey zones in Figs. 2c, S3b), and its maximum amplitude of the Hall-like signal in the Nb/SkX system is nearly two orders of magnitude larger than that of the bare Nb film.

To visually correlate these transport anomalies with the underlying topological magnetic textures, we employed cryogenic magnetic force microscopy (MFM) to image the real-space domain evolution. The low lower-critical field of the thin Nb layer allows the fringing fields of the ferromagnetic domains to be resolved directly through the superconductor. As seen in Fig. 2e, the number of negatively oriented skyrmions gradually increase as the decrease of magnetic field (A→B→C); when the direction of magnetic field reverses, the negatively oriented domains then switch into positive ones (D→E); finally, further increase of the reversed magnetic field results in the progressively diminish of the positively oriented skyrmions (F→G→H). Strikingly, a comparative analysis of the transport data and MFM images (Figs. 2d and e) reveals that the giant even-in-field $\rho_{xy}$ emerges exclusively in the superconducting transition regime where skyrmions are present. Furthermore, in stark contrast to the Nb and Nb/non-SkX references, this giant even-in field $\rho_{xy}$ in Nb/SkX varies non-monotonically with increasing temperature (7.4~8.6 K) and is confined to the existing window of non-zero topological Hall effect (THE)[43], where the LSMO layer hosts large-area skyrmion phase, as seen in Fig. S4. Together, these observations tie the anomalous transverse response to the coexistence of superconducting vortices and magnetic skyrmions.

## Skyrmion nematicity imprints anisotropic vortex dissipation

To pin down the microscopic mechanism that derives this giant even-in-field transverse Hall-like signal, we examine the scaling relationship between $\rho_{xx}$ and $\rho_{xy}$ within the superconducting transition regime. Power-law fits ($|\rho_{xy}| \propto \rho_{xx}^{K}$) yield exponents of $K = 1.03{\sim}1.19$ for temperature sweeps at fixed magnetic fields (Fig. 3a) and $K = 0.82{\sim}0.88$ for field sweeps at fixed temperatures (Fig. S5). The corresponding isothermal-field sweeps and fixed-field temperature sweeps are provided in Supplementary Figs. S4, S6–S8 of Supplementary Sections 5 and 6. These values drastically deviate from $K \approx 2$ which typifies thermally activated flux-flow Hall transport in superconductors[44]. Instead, the scaling $K \approx 1$ implies that the transverse signal is strictly proportional to $\rho_{xx}$ itself. This linear relationship is the hallmark of a resistivity-anisotropy (planar-Hall-like) origin rather than a genuine Hall channel. To rigorously test this in-plane anisotropy, we fabricated two orthogonal Hall bars aligned along the $[100]_{PC}$ and $[010]_{PC}$ crystallographic directions of LSMO on the same

Nb/SkX hybrid system. As shown in Fig. 3b, the even signal exhibits a stark polarity reversal between these orthogonal current directions at 8.4 K. Moreover, the Hall response is substantially reduced when the current is applied along the $[110]_{PC}$ direction (Fig. S9a and b). Notably, neither the bare Nb nor the Nb/non-SkX films exhibit any even-in-field Hall-like signal and in-plane anisotropy (Fig. S9c and d). The sign reversal of the field-even transverse response upon a 90° rotation of the current direction is incompatible with an isotropic in-plane transport response. This macroscopic transport anisotropy prompts us to investigate the intrinsic geometric structure and spin gradient of the skyrmions. Detailed real-space MFM imaging (Fig. 3c) reveals that the skyrmions in the bare 150-nm LSMO are inherently elliptical. The ratio of their diameters along $30°$ and $120°$ (relative to the $[100]_{PC}$ axis) is $1.25 \pm 0.25$, with their long axes consistently aligned along $30°$. The elliptical skyrmions are largely dictated by the strain-driven magnetic anisotropy and the anisotropic Dzyaloshinskii–Moriya interactions[43,45,46], establishing a robust nematic order within the skyrmion lattice.

Taken together, these observations suggest a compelling physical picture: the underlying skyrmions imprint their nematic order directly onto the superconducting vortex system. Specifically, interfacial exchange coupling between the normal-state vortex core and the anisotropic spin gradient in skyrmions renders the vortex viscosity strongly direction-dependent ($\eta_a \neq \eta_b$), the so-called nematic fluxon-skyrmion interaction. As illustrated in Fig. 3d, the anisotropic force-balance model gives the vortex velocity parallel to a current density $j$ applied at an angle $\theta$ to a principal viscosity axis (see Supplementary Information Section 10 for the full force-balance derivation):

$$v_{\parallel} = \frac{sj\Phi_0}{2}\left(\frac{1}{\eta_a} - \frac{1}{\eta_b}\right)\sin(2\theta),$$

where $s = sgn(B_z)$, and $\Phi_0$ is the flux quantum and $\eta_a$ and $\eta_b$ are the principal vortex viscosities. This compact relation captures the central physics: anisotropic drag generates the current-parallel vortex motion responsible for the field-even transverse voltage, which vanishes for $\eta_a = \eta_b$ and reverses sign under a 90° current rotation through its sin(2$\theta$) dependence, directly accounting for Figs. 2d and 3b. The full force-balance derivation, phenomenological fits, and comparison in theories are provided in Supplementary Section 10. This single framework elegantly accounts for the above giant even-in-field Hall-like signal (Fig. 2d), both the anomalous scaling exponent (Fig. 3a) and the orthogonal polarity reversal (Fig. 3b). Crucially, trivial origins for a field-even transverse voltage—such as contact misalignment, $T_c$ inhomogeneity, or a parallel conduction channel through the metallic LSMO—can be definitively ruled out: such artifacts would manifest equally in the identically fabricated Nb and Nb/non-SkX samples (Figs. S3 and S9), and fundamentally cannot produce a systematic sign reversal upon 90° rotation or respond dynamically to skyrmion density. Additionally, the reversal of polarity in the transverse response between the field-down and field-up sweeps (Fig. S10) is consistent with Bean critical-state hysteresis: reversing the sweep direction reverses the critical-state flux-density gradient and the associated shielding-current direction, thereby reconfiguring the net vortex motion (see Supplementary Note Section 10.4 for details).

Furthermore, to unequivocally distinguish interfacial exchange coupling from purely magnetostatic interactions (stray field), we performed control experiments by inserting a 2-nm insulating MgO barrier between the Nb and SkX layers. Although this ultrathin barrier leaves the stray fields virtually unperturbed, it completely suppresses interfacial exchange, thereby entirely eliminating the even transverse response (Fig. S11). This critical observation confirms that the nematic fluxon-skyrmion interaction is driven intrinsically by interlayer exchange coupling. Thus, we understand that the vortex motion in the coupled fluxon–skyrmion system can be deterministically controlled by three experimental knobs: current orientation, sweep direction, and temperature within the SkX window, and more critically, the skyrmion itself provides a locally rewritable landscape, opening a route towards spatially programmable fluxonics.

## Reconfigurable fluxon transport by skyrmion writing

This comprehensive understanding of the nematic fluxon-skyrmion interaction offers an unprecedented opportunity: the on-demand tailoring of superconducting dissipation and vortex dynamics using skyrmions as programmable control knobs at nanoscale. As schematically shown in Fig. 4a, stray field from a scanning MFM tip has been adopted to accurately write and move skyrmions in the Nb/SkX hybrid system. The detailed control of skyrmions using scanning MFM tip are provided in method section and elsewhere[30]. With such a powerful strategy, we carry out the in-situ transport measurements combining the magnetic tip writing and imaging as shown in the schematic of Fig. 4b. Remarkably, the magnitude of the anomalous Hall-like response scales linearly with the number of newly "written" skyrmions. This phenomenon provides unequivocal evidence that locally generated topological spin textures directly translates to a programmable modulation of the macroscopic vortex dissipation and fluxon transport.

Pushed to its ultimate limit, this concept can allow the density, spatial position, and even ordering arrangement of skyrmions to precisely manipulate superconducting vortex trajectories (Fig. 4c). We have also successfully used the MFM tip to precisely displace individual skyrmions within the hybrid device (Fig. 4d). Because the anisotropic drag landscape is intrinsically bound to the skyrmions themselves, relocating a skyrmion fundamentally shifts the localized friction experienced by nearby moving vortex. This reversible and real-space reconfigurability allows superconducting fluxons to be deterministically steered with a precision akin to that of an Olympic curling stone guided on ice, where locally brushing the ice surface dictates the trajectory. Consequently, this discovery establishes a conceptual shift in vortex physics from suppressing vortex motion to program it on demand, thereby converting vortex dynamics from an unavoidable and uncontrolled source of dissipation into an actively reconfigurable degree of freedom which could be a prerequisite for fluxonic quantum devices[8,9,20]. We consider that electrical transport in various cutting-edge designed superconducting devices based on our hybrid system may provide not only a microscopic probe of emergent topological states such as braidable fluxon-skyrmion excitations[39], but also selective and reversible programme of macroscopic quantum coherence for integratable quantum devices[7,8,47,48].

## Acknowledgement

This work was supported by the National Natural Science Foundation of China (NSFC Grant No. 52225205, 12550004, 12374037 and 12374100), the National K&D Program of China (Grants No. 2023YFA1406500, 2021YFA0718700 and 2024YFA1409001) and the Fundamental Research Funds for the Central Universities.

### Author contributions

J. Z., H. L., X. B., S. W. and D. S. conceived and designed the experiments. D. S. and G. L. are responsible for the thin films' fabrication. S. W. and H. L. performed theoretical analysis and discussions. X. B., D. S. and T. W. designed and fabricated Hall bars and performed the transport measurements. X. B. and T. W. performed magnetic force microscope, structural characterization and magnetic property measurements. J. Z supervised the experimental study and G. Y. provided the experimental discussion. H. L. supervised the theoretical study and K.S provided the theoretical discussion. J. Z., H. L., X. B., S. W., D. S. and Y. Z. wrote the paper.

### Competing financial interests

The authors declare no competing financial interests.

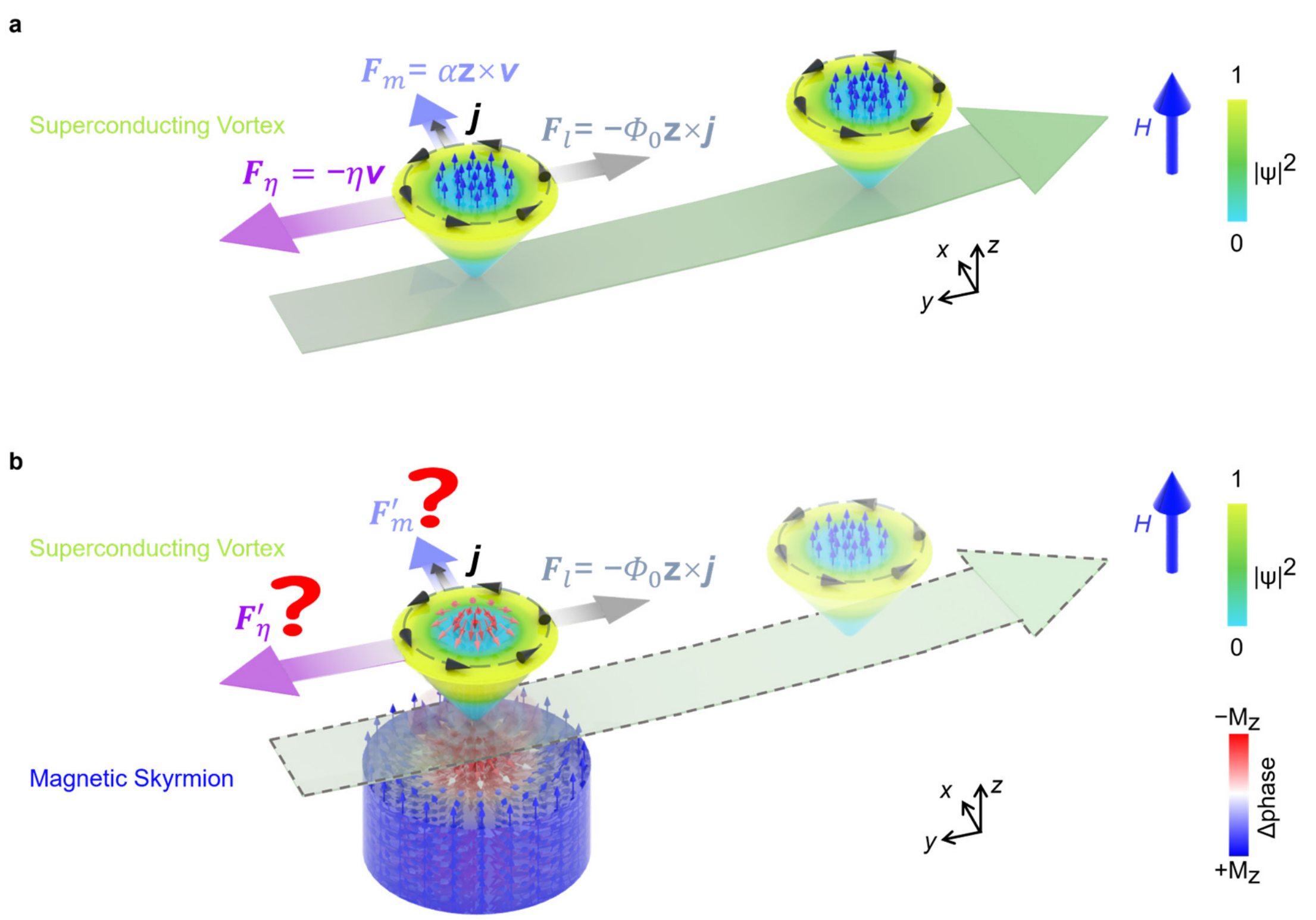

**Fig. 1 | Scenario of magnetic skyrmion control of superconducting vortex dynamics**. **a**, force balance on a moving superconducting vortex under an out-of-plane magnetic field and an applied current $\boldsymbol{j}$ (black arrow). The moving vortex experiences a Lorentz force $\boldsymbol{F}_l = -\Phi_0 \boldsymbol{z} \times \boldsymbol{j}$ (gray arrow, $\Phi_0$ is the flux quantum, $\boldsymbol{z}$ the unit vector along the vortex line), a transverse Magnus force $\boldsymbol{F}_m = \alpha \boldsymbol{z} \times \boldsymbol{v}$ (light blue arrow, $\alpha$ is the Magnus coefficient, $\boldsymbol{v}$ the vortex velocity) and a viscous drag force $\boldsymbol{F}_\eta = -\eta \boldsymbol{v}$ (purple arrow, $\eta$ is the vortex viscosity). Given that $\alpha \ll \eta$ (arrows not to scale), the force balance results in a trajectory (green arrow). **b**, when the vortex passes above a skyrmion, its normal-state core interacts with the spin gradient of skyrmion, which may modify the viscous drag force $\boldsymbol{F}_\eta{}'$ and the transverse Magnus force $\boldsymbol{F}_m{}'$, thereby altering the vortex trajectory.

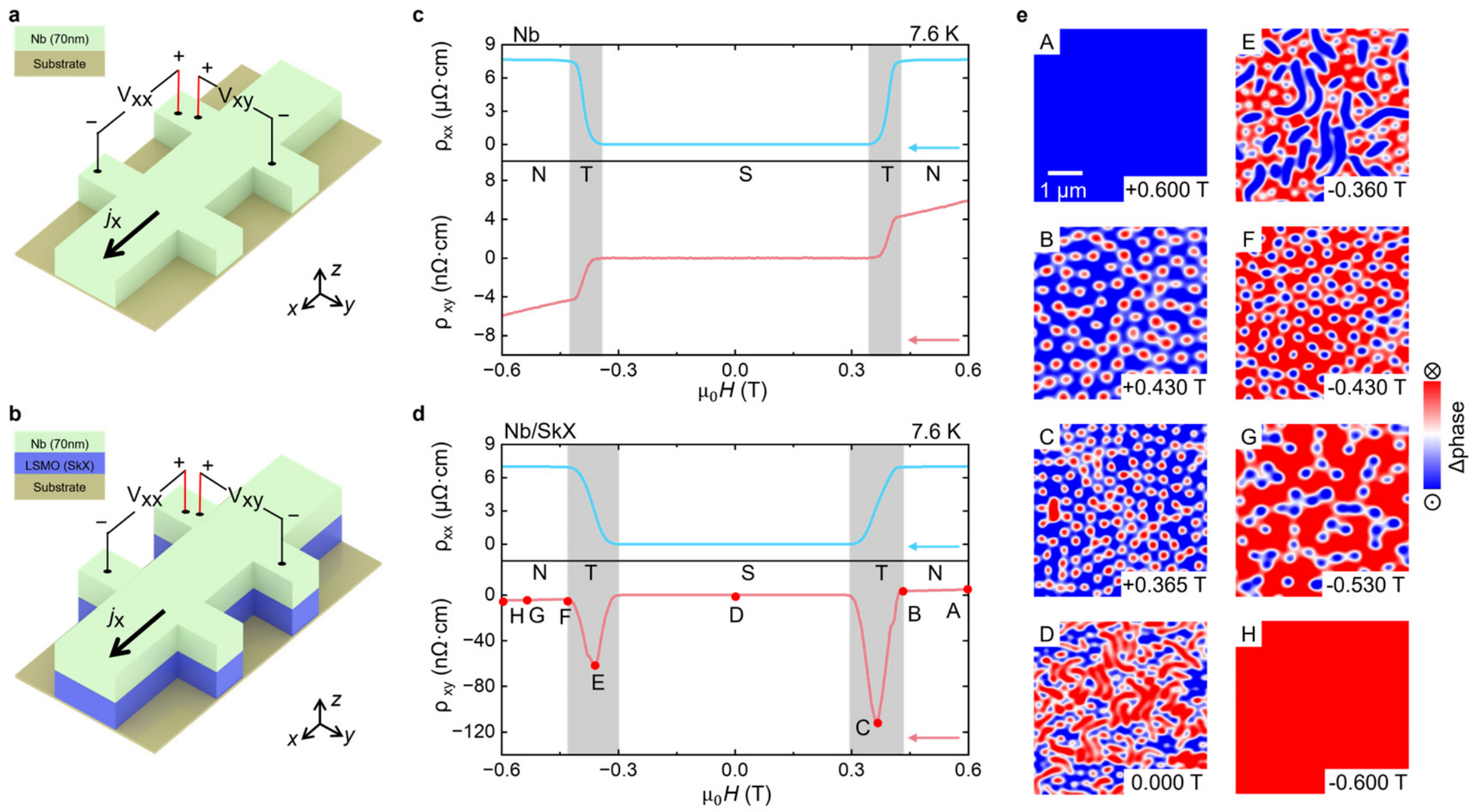

**Fig. 2 | A giant even-in-field Hall-like response in a vortex-skyrmion hybrid system. a**, **b**, schematics of transport measurement on the bare Nb (70 nm) film and the Nb/SkX hybrid system. DC current is applied along the x-axis in the film plane, and the voltages along the x-axis and y-axis are measured to obtain the longitudinal resistivity ($\rho_{xx}$) and the transverse resistivity ($\rho_{xy}$) respectively. **c**, field-down sweeps of the $\rho_{xx}$ (blue) and $\rho_{xy}$ (red) for the bare Nb film measured at 7.6 K. Arrows indicate the field-sweep direction. The gray zones mark the superconducting transition regime (T), through which the system passes from the normal state (N) to the superconducting state (S) and back to N by sweeping magnetic field. **d**, field-down sweeps of the $\rho_{xx}$ (blue) and $\rho_{xy}$ (red) for the Nb/SkX hybrid system at 7.6 K. A giant even-in-field Hall-like signal emerges within the superconducting transition regime. Its maximum amplitude in the Nb/SkX system is nearly two orders of magnitude larger than that in the bare Nb film. **e**, magnetic force microscopy images acquired from the Nb/SkX hybrid system at the fields corresponding to points A to H marked in d.

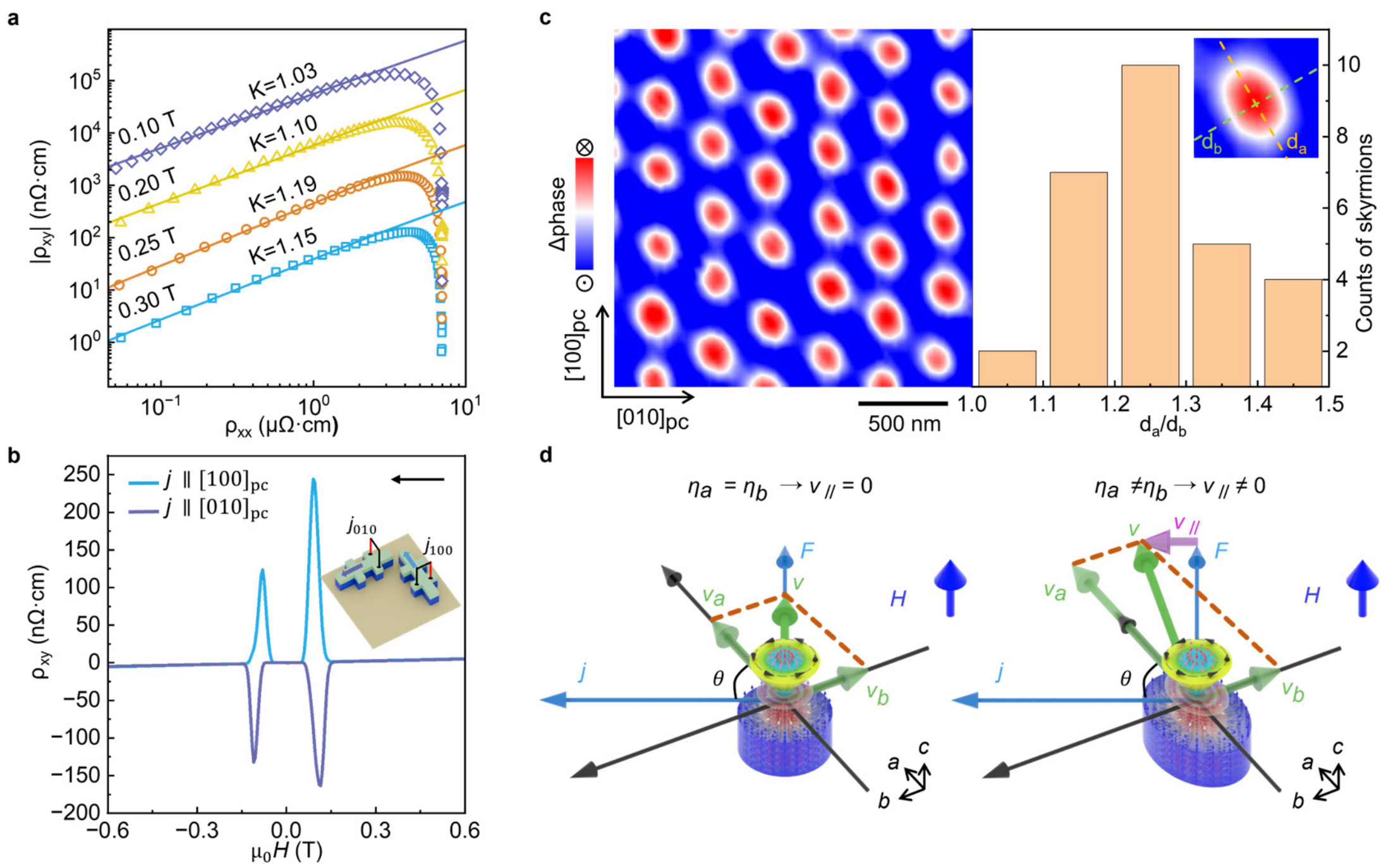

**Fig. 3 | Nematic origin of the giant even-in-field Hall-like response. a**, $|\rho_{xy}|$ versus $\rho_{xx}$ on a log–log scale at fixed magnetic fields of 0.10–0.30 T. Power-law fits $|\rho_{xy}| \propto \rho_{xx}^{K}$ yield $K = 1.03{\sim}1.19$, far from the $K \approx 2$ which is conventional flux-flow Hall transport in superconductors. **b**, $\rho_{xy}$ (H) on the field-down sweep at 8.4 K for two orthogonal Hall bars on the same film, with current along $[100]_{PC}$ (blue) and $[010]_{PC}$ (purple) of the pseudocubic LSMO lattice: the even signal reverses polarity with current directions. **c**, magnetic force microscopy image and diameter statistics of skyrmions along $30°$ ($d_a$) and $120°$ ($d_b$) with respect to $[100]_{PC}$. The ellipticity $d_a/d_b = 1.25 \pm 0.25$ with the same direction of long axis reveals nematic order of the skyrmion shapes. Scale bar, 500 nm. **d**, with isotropic drag ($\eta_a = \eta_b$, left) in skyrmion with isotropic shape, vortex motion yields only the conventional field-odd Hall-like response; with anisotropic drag ($\eta_a \neq \eta_b$, right) in skymion with anisotropic shape, the vortex gains a velocity component along the current (red dashed trajectory), generating a giant transverse signal that is even in field and reverses sign under a $90°$ rotation of the current.

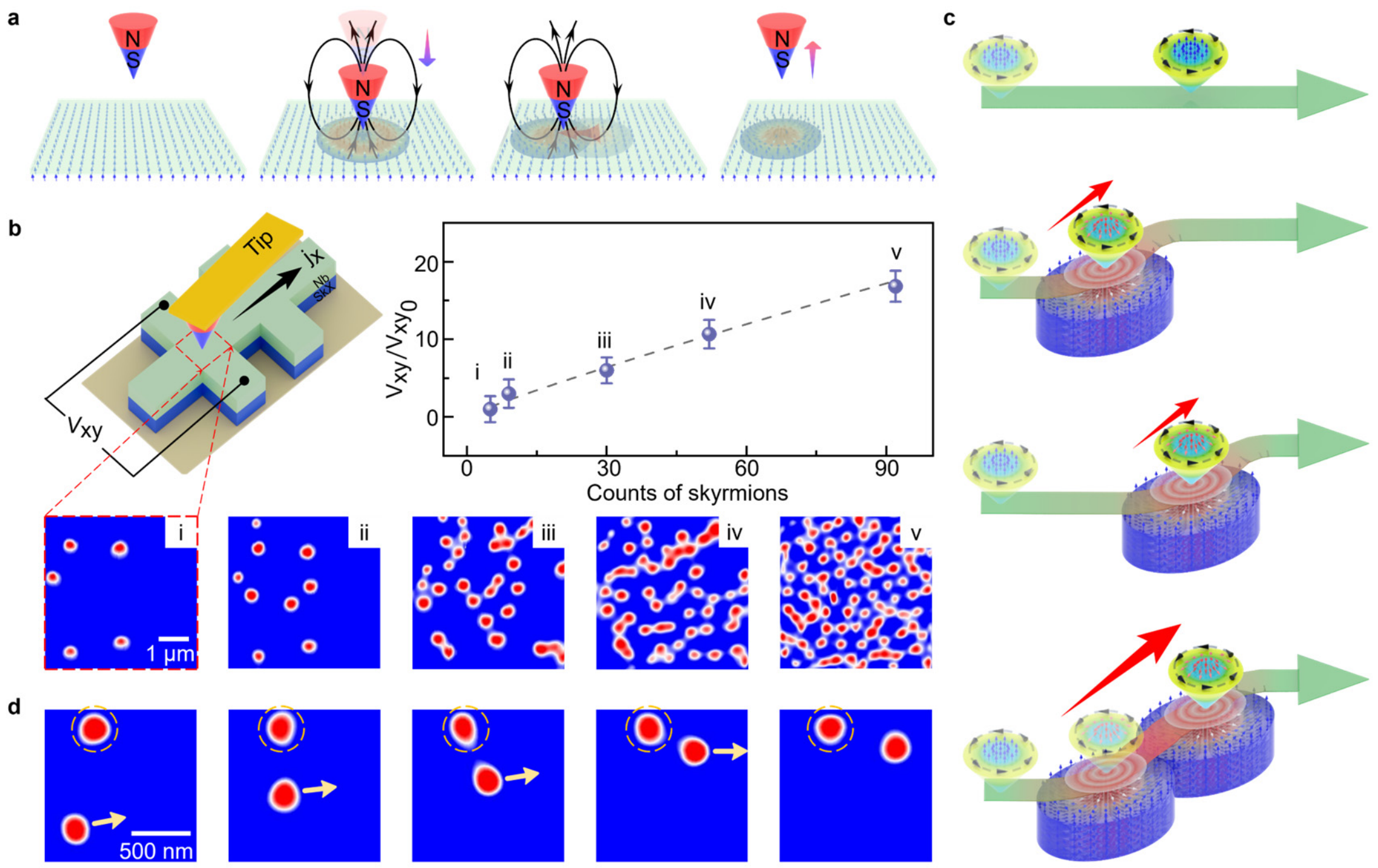

**Fig. 4 | Nanoscale manipulation of vortex dissipation and fluxon transport by skyrmion writing and moving. a**, schematic illustration of a single skyrmion written and moved by a magnetic tip. **b**, transport measurement with in-situ writing of skyrmions using a magnetic tip on the Nb/SkX hybrid device at fixed temperature and magnetic field. The normalized Hall signal increases monotonically with the number of skyrmions; the corresponding magnetic force microscopy images track the count variation. **c**, under an applied current, the trajectory of a superconducting vortex is determined by the balance of the Lorentz force, the Magnus force, and the viscous drag. Upon writing a skyrmion, the vortex velocity is redirected, resulting in an altered trajectory (red trajectory). Moreover, by moving the skyrmion, the deflection of the vortex motion and the accompanying fluxon transport can be controlled at different spatial locations. **d**, pushing forward a skyrmion in the Nb/SkX hybrid device using a magnetic tip.